\documentclass[journal,twocolumn,singlespaced]{IEEEtran}

\usepackage[utf8]{inputenc}
\usepackage{amsmath, amssymb, amsfonts, amsbsy, latexsym, amsthm}
\usepackage[english]{babel}
\usepackage{doi}
\usepackage[normalem]{ulem}
\usepackage{lineno}
\usepackage{setspace}
\usepackage{bm}
\ifCLASSINFOpdf

\else

\fi

\usepackage{color}
\usepackage{graphicx}
\usepackage{epstopdf}
\usepackage{subfigure}
\usepackage{grffile} 
\usepackage{multirow}
\usepackage{float}

\usepackage{marginnote}

\usepackage{url}
\usepackage{hyperref}
\hypersetup{
colorlinks,%
citecolor=blue,%
filecolor=blue,%
linkcolor=blue,%
urlcolor=blue
}

\renewcommand{\baselinestretch}{1.0}
\begin{document}
\title{CNN-tuned spatial filters for P- and S-wave decomposition and applications in elastic imaging}

\author{
\IEEEauthorblockN{Wenlong~Wang\IEEEauthorrefmark{3}, Jianwei Ma\IEEEauthorrefmark{4}, George A. McMechan\IEEEauthorrefmark{6} }\\
\thanks{
\IEEEauthorblockA{\IEEEauthorrefmark{3}Center of Geophysics, Department of Mathematics and Artificial Intelligence Laboratory, Harbin Institute of Technology, Harbin, China 150001 (e-mail:wenlong.wang@hit.edu.cn).}
\IEEEauthorblockA{\IEEEauthorrefmark{4}School of Earth and Space Sciences, Peking University, Beijing, China, 100871.}
\IEEEauthorblockA{\IEEEauthorrefmark{6}Center for Lithospheric Studies, The University of Texas at Dallas, 800 W Campbell Road (ROC21), Richardson, TX, USA 75080.}
}
}

\maketitle

\begin{spacing}{1.0}
\begin{abstract}
P- and S-wave decomposition is essential for imaging multi-component seismic data in elastic media.
A data-driven workflow is proposed to obtain a set of spatial filters that are highly accurate and artifact-free in decomposing the P- and S-waves in 2D isotropic elastic wavefields. The filters are formulated initially by inverse Fourier transforms of the wavenumber-domain operators, and then are tuned in a convolutional neural network to improve accuracy using synthetic snapshots.
Spatial filters are flexible for decomposing P-and S-waves at any time step without performing Fourier transforms over the entire wavefield snapshots, and thus are suitable for target-oriented imaging.
Snapshots from synthetic data show that the network-tuned spatial filters can decompose P- and S-waves with improved decomposition accuracy compared with other space-domain PS decomposition methods.
Elastic reverse-time migration using P- and S-waves decomposed from the proposed algorithm shows reduced artifacts in the presence of a high velocity contrast.

\end{abstract}


\section{Introduction}\label{section:introduction}

\IEEEPARstart{P}- and S-waves coexist in seismic waves propagating through the solid earth.
The causes, polarization directions, and propagation velocities, of P- and S-waves are different, and thus they carry different information from the subsurface at a variety of scales. For example, in seismic imaging for oil and gas exploration, P- and S-wave reflections can be imaged independently. In the presence of gas clouds and fractures, the P-waves are strongly attenuated, while S-waves are less affected, and S-wave images provide valuable structural information from reservoirs \cite{li98, knapp01}.
In computational seismology, most imaging methods for earthquake and microseismic events involve picking the P- and S-wave arrival times. Artman \emph{et al.} \cite{artman10} and Yang and Zhu \cite{jidong18} propose time-reversal imaging of earthquake ruptures and hydraulic fractures by separating and propagating the recorded P- and S-waves back into the earth, and applying the elastic imaging conditions. All of these applications require separating the P- and S-waves accurately and efficiently.

One approach to separate and utilize both P- and S-wave data is the ray-based method. Examples include Kirchhoff migration \cite{kuo84,dai86,hokstad00}, in which  the PS separation is implicit by using P- and S-velocity models separately for ray-tracing to compute the image times. Ray-based methods are efficient, but their high frequency assumption makes ray theory unable to fully describe wave phenomena, especially for complicated models \cite{gray01}.
Another category is wave-equation based solutions \cite{chang86,chang94,whitmore95}, which reconstruct the full elastic receiver wavefields from boundary conditions \cite{wapenaar90}, and separates the P- and S-waves before, or as a part of, application of the image condition \cite{yan09}.

The divergence and curl operators, which are based on Helmholtz decomposition \cite{aki80}, are widely applied in PS separations. However, the phases and amplitudes of the separated P- and S-waves are distorted by the divergence and curl operators, causing problems for true amplitude imaging and subsequent interpretations \cite{sun01}. To solve this problem, PS wavefield decomposition is proposed, which preserves the vector components of the input elastic wavefield \cite{ma03,zhang10,wenlong_cmp15}.

There are two groups of algorithms for PS decomposition depending on the domain of calculation. In the wavenumber domain,
elastic wavefields are projected to the P- and S-polarization directions \cite{zhang10}; Zhu \cite{zhu17} decomposes the wavefields by solving a Poisson's equation. Both of these methods involve Fourier transforms. However, it is expensive, or even prohibitive, to perform Fourier transforms on large (3D) datasets. Xiao and Leaney \cite{xiao10} introduce an auxiliary stress wavefield into the elastodynamic wave equations to obtain decoupled wavefields, but the decomposition is embedded in the elastic extrapolations (decoupled propagation), and artifacts are likely to be generated at large velocity contrasts in the model \cite{wenlong_cmp15}. Yan and Sava \cite{yan09} propose spatial filters for separating P- and S-waves in transversely isotropic media with a vertical axis (VTI). The spatial filters don't require Fourier transforms, but the accuracy is limited by the truncation of the spatial filters. 

PS separation and decomposition can also be achieved with deep learning algorithms. Wei et al. \cite{wei19} reconstruct P-waves in VSP data using conditional generative adversarial network. Wei et al. \cite{wei20} extend the algorithm to separate P- and S-waves in multi-component VSP data using multi-task learning. However, these methods need a lot of training samples from a variety of velocity models to compensate the lack of physical information.
The spatial filters are convolutional operators, which are widely used in the computer vision community [i.e. convolutional neural networks (CNNs)] \cite{lecun89}. Thus, CNNs are suited for separating P- and S-wave snapshots.
Kaur \emph{et al.} \cite{kaur19} apply a generative adversarial network (GAN) to match the predicted P-waves with the decomposed ones from low-rank approximation in anisotropic media; the network needs to be re-trained to apply to different models. 
We propose to formulate the PS decomposition as a single-layer CNN with the coupled wavefield snapshots as inputs and the decomposed P-wave snapshots as outputs. 
The spatial filters are tuned to optimize PS decomposition accuracy by training in a training set. Different from the existing methods, our network only needs to be trained once on a complex model and can be applied to any models, because the physics in PS decomposition are embedded in the network structure.
The trained network can decompose elastic wavefields in different models without re-training. 

This paper is organized as follows, we first illustrate the wavenumber domain operators for P- and S-wave decomposition, and their relation to spatial filters. Then we describe a CNN structure to tune the spatial filters. 
Wavefield snapshot decomposition and elastic reverse time migration are performed to document the improvements in accuracy by comparing the decomposition results with those from
two other space-domain PS decomposition methods. To limit the scope, we focus on 2D isotropic media, although PS decompositions in anisotropic and anelastic media are also possible \cite{cheng14, wenlong_visco15, wenlong18}.

\section{PS decomposition}
Different from the commonly used divergence and curl operators \cite{aki80} which distort the amplitudes and phases during wavefield separation, PS decomposition preserves all the components of the elastic vector wavefield. In the wavenumber domain, the explicit equations to obtain 2D decomposed P-waves are \cite{zhang10}
\begin{equation}
\tilde{U}^P_x = K^2_x \tilde{U}_x + K_x K_z \tilde{U}_z,
\label{eqn:p_wave1}
\end{equation}
and
\begin{equation}
\tilde{U}^P_z = K^2_z \tilde{U}_z + K_z K_x \tilde{U}_x,
\label{eqn:p_wave2}
\end{equation}
where $K_x$ and $K_z$ are the components of the normalized wavenumber vector $\boldsymbol{K}$, which represents the P-wave polarization direction in isotropic media. $U_x$ and $U_z$ are the space-domain elastic wavefield components with coupled P- and S-waves, and a tilde over the wavefield indicates the wavenumber-domain representation.
The S-waves can be decomposed in a similar way, or by subtracting the P-waves from the coupled wavefields, component-by-component-by
\begin{equation}
\tilde{U}^S_x = \tilde{U}_x - \tilde{U}^P_x,
\label{eqn:s_wave3}
\end{equation}
and
\begin{equation}
\tilde{U}^S_z = \tilde{U}_z - \tilde{U}^P_z,
\label{eqn:s_wave4}
\end{equation}
in either the space or the wavenumber domain. Thus only the P-wave decomposition is discussed and demonstrated in the following parts of this paper.

\section{Spatial filters for P- and S-wave decomposition}\label{sec1}
\begin{figure}[!t]
  \centering
  \includegraphics[width=3.4in]{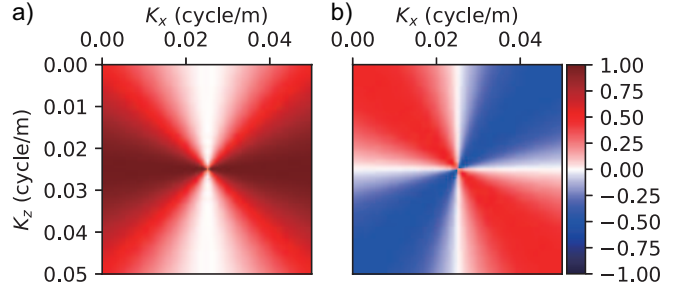}
  \caption{The wavenumber domain operators (a) $K^2_x$ and (b) $K_xK_z$ for isotropic P- and S-wave decompositions.}
  \label{fig:K_filter2.eps}
\end{figure}

\begin{figure}[!t]
  \centering
  \includegraphics[width=3.4in]{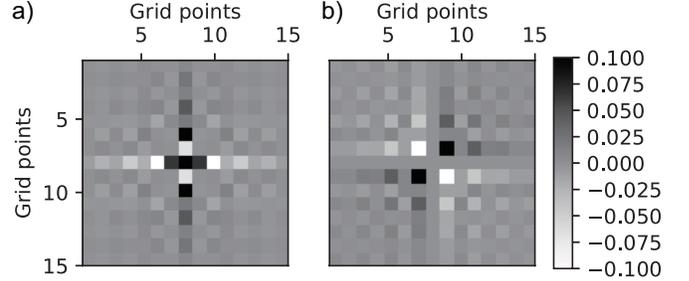}
  \caption{The (a) $L_x$ and (b) $L_{xz}$ spatial filters, which are transformed from the wavenumber domain operators $K^2_x$ and $K_xK_z$ in Figure~\ref{fig:K_filter2.eps}. $L_z$ is the transpose of $L_x$.}
  \label{fig:filter.eps}
\end{figure}

In the 2D equations~\ref{eqn:p_wave1} and \ref{eqn:p_wave2}, three wavenumber operators $K^2_x$, $K^2_z$ and $K_xK_z$ are involved; the $K^2_z$ is the transpose of $K^2_x$ when the spatial intervals along $x$ and $z$ directions are the same. $K^2_x$ and $K_xK_z$ are plotted in Figures~\ref{fig:K_filter2.eps}a-\ref{fig:K_filter2.eps}b, respectively.
The wavenumber domain operators can be transformed into the space-domain by inverse Fourier transforms to formulate the spatial filters \cite{yan09}. The equivalent expressions of equations~\ref{eqn:p_wave1} and \ref{eqn:p_wave2} in the space-domain are
\begin{equation}
U^P_x = L_x [U_x] + L_{xz} [U_z],
\label{eqn:p_space1}
\end{equation}
and
\begin{equation}
U^P_z = L_z [U_z] + L_{xz} [U_x],
\label{eqn:p_space2}
\end{equation}
where $L_x$, $L_z$ and $L_{xz}$ are filters that are designed to decompose the elastic wavefields.
The square brackets represent spatial filtering of the wavefields.
Different from the anisotropic media, the polarizations of P- and S-waves in isotropic wavefield have much simpler relations with the wavenumbers, they do not depend on the model parameters ($V_P$, $V_S$ or $\rho$). Thus the spatial filters are stationary and universal for different models. The spatial filters are trained only once using synthetic labelled data pairs, and the trained filters can be applied to decompose P- and S-waves in any isotropic wavefields.
Full representations of the wavenumber domain operators are infinitely large 2D patches in the space-domain which make them infeasible to be applied. Yan and Sava \cite{yan09} propose to truncate the filters to balance the computation efficiency and the separation accuracy.
A set of truncated $L_x$ and $L_{xz}$ filters of size 15 $\times$ 15 are shown in Figures~\ref{fig:filter.eps}a-\ref{fig:filter.eps}b ($L_z$ is the transpose of $L_x$, and thus is not plotted).

\section{Tuning the spatial filters with CNN}
Truncations of the filters reduce the accuracy and completeness of the decomposition. To improve the accuracy without increasing the size of the filters, we propose to tune the spatial filters $L_x$ and $L_{xz}$ by training in a CNN.
Following equations~\ref{eqn:p_space1} and \ref{eqn:p_space2}, a CNN is constructed to obtain the P-waves $\boldsymbol{U}^P$ from the coupled wavefield snapshot $\boldsymbol{U}$ of the form
\begin{equation}
\boldsymbol{U}^P = \mathrm{CNN}(\boldsymbol{U}).
\label{eqn:cnn}
\end{equation}
The two components ($U_x$ and $U_z$) of the elastic wavefield snapshot $\boldsymbol{U}$ are input to the network as two channels, and the output is expected to be the decomposed 2-component P-wave snapshot $\boldsymbol{U}^P$.

\begin{figure}[!t]
  \centering
  \includegraphics[width=3.2in]{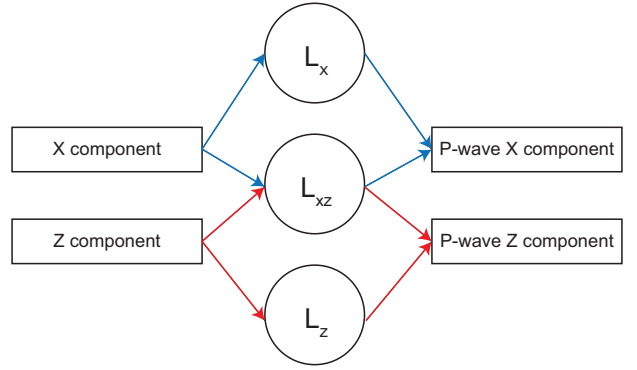}
  \caption{
The CNN architecture with respect to equations~\ref{eqn:p_space1} and \ref{eqn:p_space2} for tuning the spatial filters. The arrows indicate the data flow direction, and the $L_x$, $L_z$ and $L_{xz}$ are convolutional operators that operate on the input wavefields. The $L_z$ can be obtained by transposing $L_x$, and thus they share the same filter values during training.
}
  \label{fig:cnn.eps}
\end{figure}

The CNN for PS decomposition is shown in Figure~\ref{fig:cnn.eps}. Two spatial filters ($L_x$ and $L_{xz}$) are to be trained.
Most neural networks in deep learning are black-boxes and cannot be explained. However, the CNN which we use contains only one CNN layer, which can be simply interpreted as spatial filters. The physical meaning of our CNN is a spatial representation of a Fourier domain operator, which is similar to the work of Yan and Sava \cite{yan09}.
Thus our network does not contain pooling or non-linear activation functions because of the well understood physics that is involved.
Another difference is that, instead of initializing the spatial filters with random numbers, they are obtained by transforming the wavenumber domain operators in Figure~\ref{fig:filter.eps}a and \ref{fig:filter.eps}b, and are set as initial values before training.

The network is trained by minimizing the loss function
\begin{equation}
L = \sum_{\substack{i=1,N}} |\boldsymbol{U}^P_{i}-\hat{\boldsymbol{U}}^P_{i}|^2,
\label{eqn:loss}
\end{equation}
where $\boldsymbol{U}^P$ and $\hat{\boldsymbol{U}}^P$ are the predicted and ground truth P-waves, respectively. $|\cdot|$ calculates the modulus arguments of vector difference. $N$ is the number of grid points in a wavefield snapshot.
The filters are tuned in a CNN training process. We expect the training to reduce the errors that are associated with the filter truncations.

\begin{figure*}[!t]
  \centering
  \includegraphics[width=6.2in]{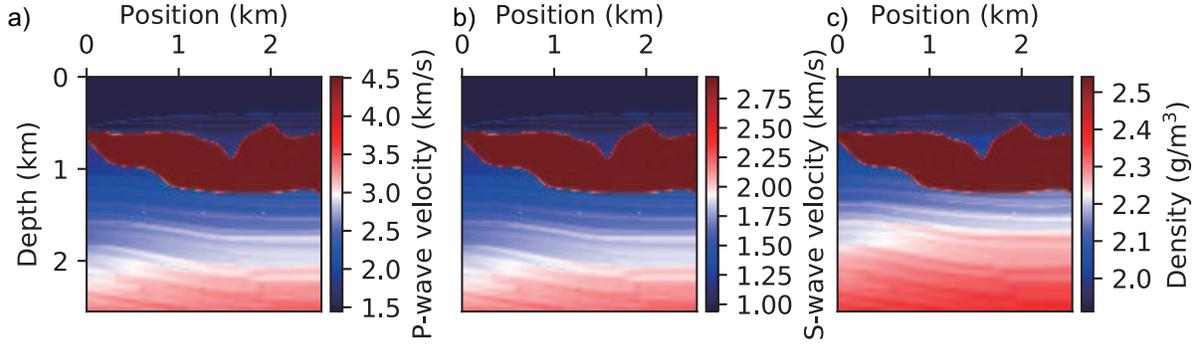}
  \caption{
The (a) Vp, (b) Vs and (c) density of the Sigsbee model for generating wavefield snapshots for training.
}
  \label{fig:sigsbee_model.eps}
\end{figure*}

We generate a training set with 800 samples, which contains coupled and decomposed P-wave snapshot pairs. The coupled wavefield snapshots are generated by simulating wave propagation in a portion of the Sigsbee model \cite{paffenholz02} (Figure~\ref{fig:sigsbee_model.eps}) using an eighth-order in space, second-order in time, stress-particle-velocity, staggered-grid, finite-difference solution \cite{virieux86}. An explosive source with a 10 Hz Ricker wavelet is placed at (x, z) = (1.28, 0.02) km. The spatial intervals in both x and z directions are 10 m, and the time interval is 1 ms. The 800 2-component snapshots used are captured at random time steps from T = 0.0 s to T = 1.4 s.
Although reflections from model edges can also be separated into P- and S-waves, we use Fourier transforms to generate the ground truth (decomposed P- and S-wave snapthots), so they may have wrap-around errors at the model edges. Thus, convolutional perfectly matched layer (CPML) absorbing boundary conditions \cite{komatitsch07} are used on all four boundaries to reduce unwanted reflections.

The training doesn't need a massive training set, because (1) approximate spatial filters (Figures~\ref{fig:filter.eps}a and \ref{fig:filter.eps}b) are set as initial filters which are close to the optimized filters; (2) the CNN structure in Figure~\ref{fig:cnn.eps} imposes a physical contraint on the training; (3) only 2 filters ($L_x$ and $L_{xz}$) need to be tuned. After training, the CNN-tuned spatial filters can be applied to any wavefield snapshot that has the same spatial intervals as the training set.

\begin{figure}[!t]
  \centering
  \includegraphics[width=3.2in]{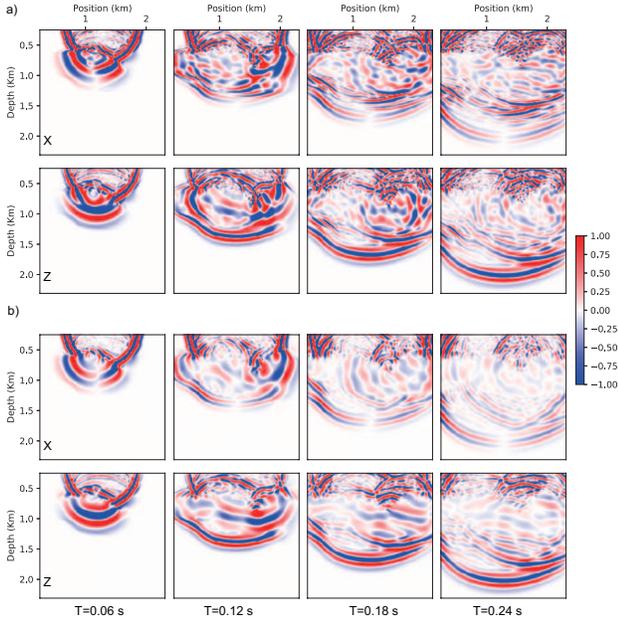}
  \caption{
Four representative snapshot pairs in the training set. (a) The coupled X and Z particle velocity components and (b) the corresponding (ground truth) decomposed P-wave x and z components obtained using equations~\ref{eqn:p_wave1}-\ref{eqn:p_wave2}.
}
  \label{fig:sample_train.eps}
\end{figure}

The ground truth P-waves are the decomposition results using the wavenumber decomposition algorithm (equations~\ref{eqn:p_wave1} and \ref{eqn:p_wave2}), which is, by far, the most accurate algorithm for P- and S-wave decomposition \cite{zhang10}. Samples of the training set are plotted in Figure~\ref{fig:sample_train.eps}. The training set is sufficiently diverse, as the Sigsbee model contains a wide range of structures and velocities.

\begin{figure}[!t]
  \centering
  \includegraphics[width=3.2in]{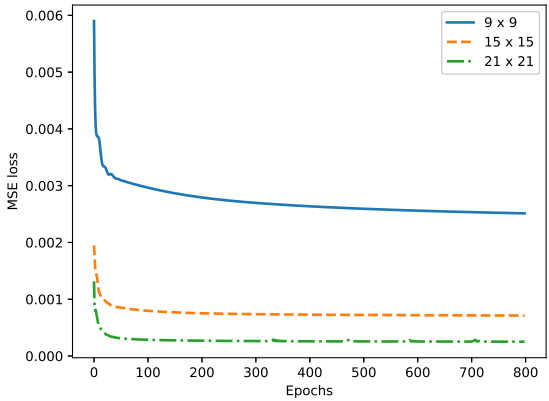}
  \caption{
The loss evolutions of different filter sizes during traning.
}
  \label{fig:cost_cmp.eps}
\end{figure}

\begin{figure}[!t]
  \centering
  \includegraphics[width=3.2in]{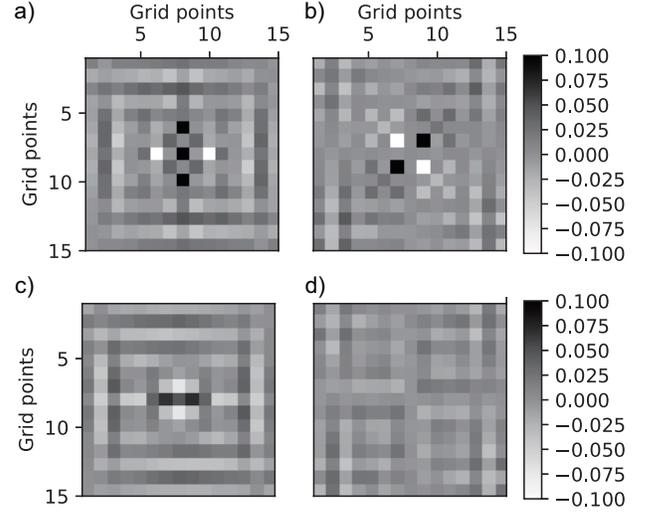}
  \caption{
The (a) $L_x$ and (b) $L_{xz}$ spatial filters after tuning with CNN. (c) and (d) show the difference between the $L_x$ and $L_{xz}$ filters before and after tuning, respectively.
}
  \label{fig:filter_tuned.eps}
\end{figure}

The size of the spatial filter should be at least one wavelength to maximize the resolution, but large spatial filters increase the computation cost.  Three sets of spatial filters are trained with dimension 9 $\times$ 9, 15 $\times$ 15 and 21 $\times$ 21.
The networks are trained using an Adam optimizer \cite{diederik14} in the Pytorch (\url{www.pytorch.org}) platform. Figure~\ref{fig:cost_cmp.eps} shows that losses decrease as the training proceeds, and that the decomposition accuracy of all these sizes of filters is improved during training. As expected, spatial filters with larger sizes can reach lower losses. Big improvements can be made when the filters expand from 9 $\times$ 9 to 15 $\times$ 15, but improvements are less significant for filter size 21 $\times$ 21 (Figure~\ref{fig:cost_cmp.eps}).
In the following tests, only the updated $L_x$ and $L_{xz}$ filters of the dimension 15 $\times$ 15 are used, which are plotted in Figures~\ref{fig:filter_tuned.eps}a and \ref{fig:filter_tuned.eps}b, respectively. Figures~\ref{fig:filter_tuned.eps}c and d show the differences between the filters before and after the tuning. 

\section{Synthetic tests and comparisons}\label{sec2}
We perform tests on synthetic data using the CNN-tuned spatial filters (CNN-SF), and the results are compared with the decomposed P-waves from untuned spatial filters (SF), and those from decoupled propagation (DP) \cite{ma03}. DP involves solving an auxiliary equation along with the elastodynamic wave equations, and thus has to be applied at each time step from the beginning of the elastic wavefield extrapolation.  Note that CNN-SF is tuned only from the Sigsbee model in the previous section, and can be directly applied to the following (different) models without re-training.

\subsection{Two-layer example}
\begin{figure}[!t]
  \centering
  \includegraphics[width=3.2in]{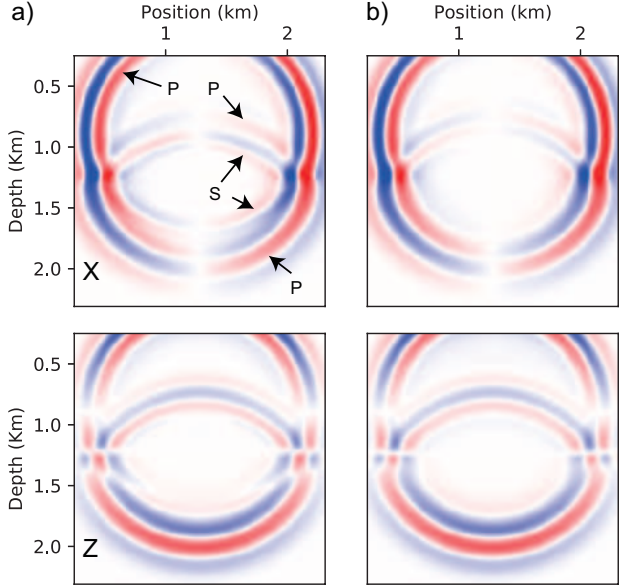}
  \caption{
The (a) coupled x and z wavefield components and (b) decomposed P-wave x and z components (the ground truth) in the two-layer model at t = 0.42 s.
}
  \label{fig:layer_snap.eps}
\end{figure}

The first test is performed on an isotropic two-layer model with $V_P$ = 3 km/s, $V_S$ =2.1 km/s, and $\rho$ = 2.2 $\mathrm{g/cm^3}$ for the upper layer, and $V_P$ = 4 km/s, $V_S$ =2.4 km/s, and $\rho$ = 2.4 $\mathrm{g/cm^3}$ for the lower layer.
The model has 10 m grid spacing in both the x- and z-directions. An explosive source with a 10 Hz Ricker wavelet is placed at (x, z) = (1.28, 0.9) km. A snapshot of the PS coupled x and z particle velocity components at t = 0.42 s is shown in Figure~\ref{fig:layer_snap.eps}a, and the ground truth P-wave using equations~\ref{eqn:p_wave1} and \ref{eqn:p_wave2} are shown in Figure~\ref{fig:layer_snap.eps}b as benchmarks.

\begin{figure}[!t]
  \centering
  \includegraphics[width=3.2in]{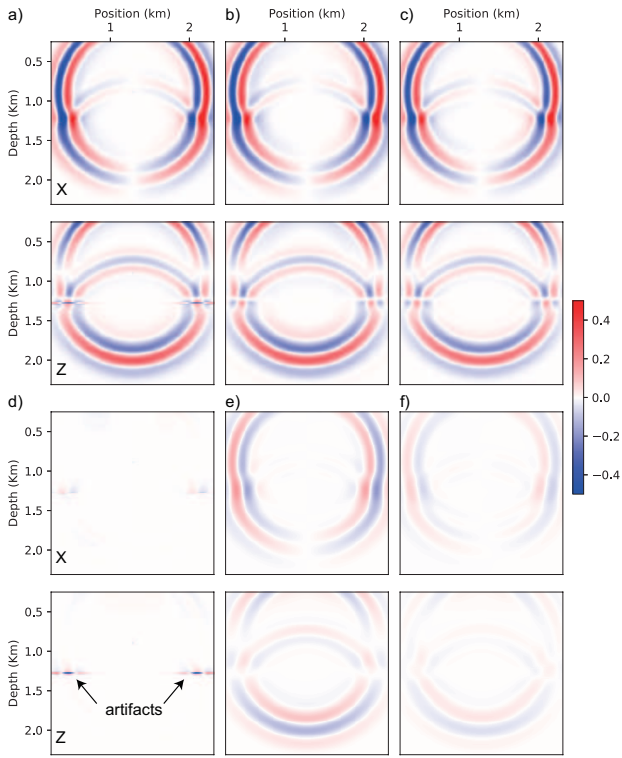}
  \caption{
The PS decomposition results using (a) DP, (b) SF and (c) CNN-SF in the two-layer model.
The decomposition accuracy can be measured by subtracting their respective decomposition results from the ground truth (Figure~\ref{fig:layer_snap.eps}b); the residuals are shown in (d), (e) and (f), respectively.
}
  \label{fig:layer_cmp.eps}
\end{figure}

The decomposed P-waves using DP, SF and CNN-SF methods are shown in Figures~\ref{fig:layer_cmp.eps}a-\ref{fig:layer_cmp.eps}c, respectively. To measure the decomposition accuracy, we subtract the decomposition results ($\boldsymbol{U}^P_{i}$) in Figures~\ref{fig:layer_cmp.eps}a-\ref{fig:layer_cmp.eps}c from the ground truth ($\hat{\boldsymbol{U}}^P_{i}$) P-wave (Figure~\ref{fig:layer_snap.eps}b), and the residuals are shown in Figures~\ref{fig:layer_cmp.eps}d-\ref{fig:layer_cmp.eps}f. DP has high accuracy at homogeneous parts of the medium, but it generates serious artifacts along the interface between layers, which are caused by the added auxiliary equation \cite{wenlong_cmp15}, indicating that the DP method can be applied only to smooth models. The spatial filter algorithms SF and CNN-SF don't have this problem. The residuals in Figure~\ref{fig:layer_cmp.eps}f are smaller than in Figure~\ref{fig:layer_cmp.eps}e, indicating that the CNN-SF achieves higher accuracy than the SF.
We quantify the accuracy by
\begin{equation}
A = 1- \frac{\sum_{\substack{i=1,N}} |\boldsymbol{U}^P_{i}-\hat{\boldsymbol{U}}^P_{i}|^2}
         {\sum_{\substack{i=1,N}} |\boldsymbol{U}^P_{i}|^2}.
\end{equation}
The calculated accuracies of DP, SF and CNN-SF in this example are 92.2\%, 90.1\%, and 98.6\%, respectively.

The computation times of DP, SF, and CNN-SF are 0.01, 0.25 and 0.25 seconds, respectively, for each decomposition of the above model. DP has the best efficiency, but it involves solving an auxiliary equation along with solving the wave equations, thus decomposition with DP is not independent from wavefield extrapolation.
CNN-SF and SF algorithms, on the other hand, don't require extrapolations, and they have the same filter size, and thus share the same computation time.

\subsection{Marmousi-2 model example}
\begin{figure*}[!t]
  \centering
  \includegraphics[width=6.2in]{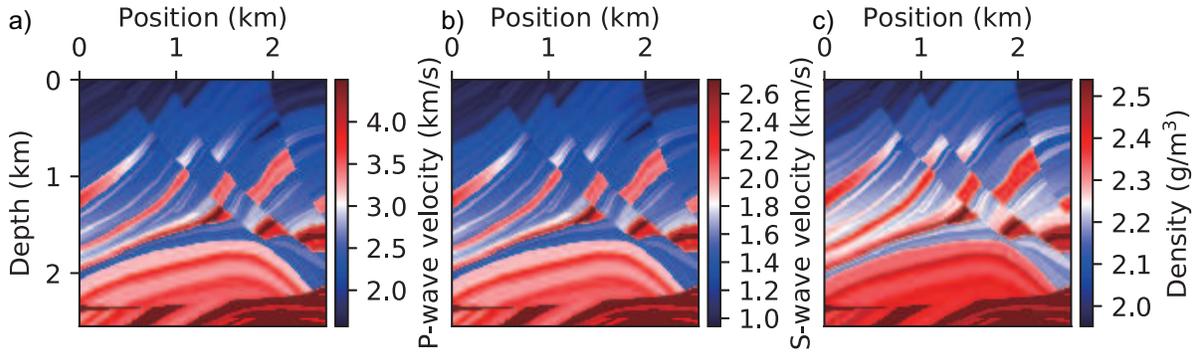}
  \caption{
The (a) P-wave velocity, (b) S-wave velocity and (c) density of the Marmousi-2 model.
}
  \label{fig:marmousi_model.eps}
\end{figure*}

\begin{figure}[!t]
  \centering
  \includegraphics[width=3.2in]{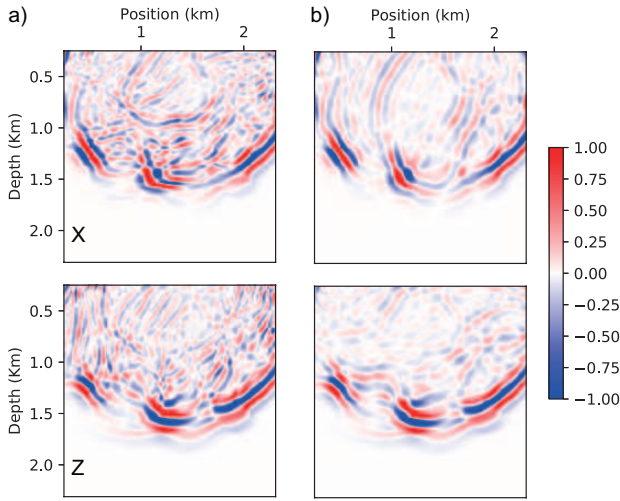}
  \caption{
The (a) coupled wavefield snapshot X and Z components and (b) decomposed P-wave snapshot X and Z components (ground truth) in the Marmousi-2 model at t = 0.21 s.
}
  \label{fig:marmousi_snap.eps}
\end{figure}

The second test is performed on a portion of the Marmousi-2 model \cite{martin02} (Figure~\ref{fig:marmousi_model.eps}).
The model has 10 m grid spacing in both x- and z-directions. An explosive source with a 15 Hz Ricker wavelet is placed at (x, z) = (1.28, 0.04) km. A snapshot of the x and z particle velocity components at t = 0.21 s is shown in Figure~\ref{fig:marmousi_snap.eps}.

\begin{figure}[!t]
  \centering
  \includegraphics[width=3.2in]{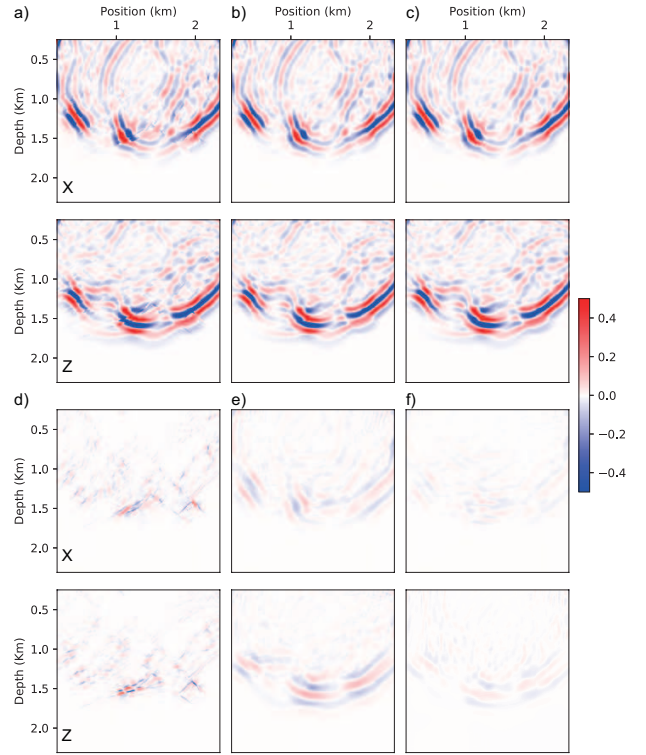}
  \caption{
The PS decomposition results using (a) DP, (b) SF and (c) CNN-SF in the Marmousi-2 model. The decomposition residuals are shown in (d), (e) and (f), respectively.
}
  \label{fig:marmousi_cmp.eps}
\end{figure}

We decompose the representative coupled wavefields in Figure~\ref{fig:marmousi_snap.eps} using the DP, SF and CNN-SF methods, respectively; the decomposed P-waves are shown in Figures~\ref{fig:marmousi_cmp.eps}a-\ref{fig:marmousi_cmp.eps}c, and their residuals are in Figures~\ref{fig:marmousi_cmp.eps}d-f, respectively. Results of DP contain artifacts along the velocity interfaces. The decomposition results from SF and CNN-SF methods don't have such artifacts, and the decomposition accuracy of CNN-SF is better than SF.

\begin{figure}[!t]
  \centering
  \includegraphics[width=3.2in]{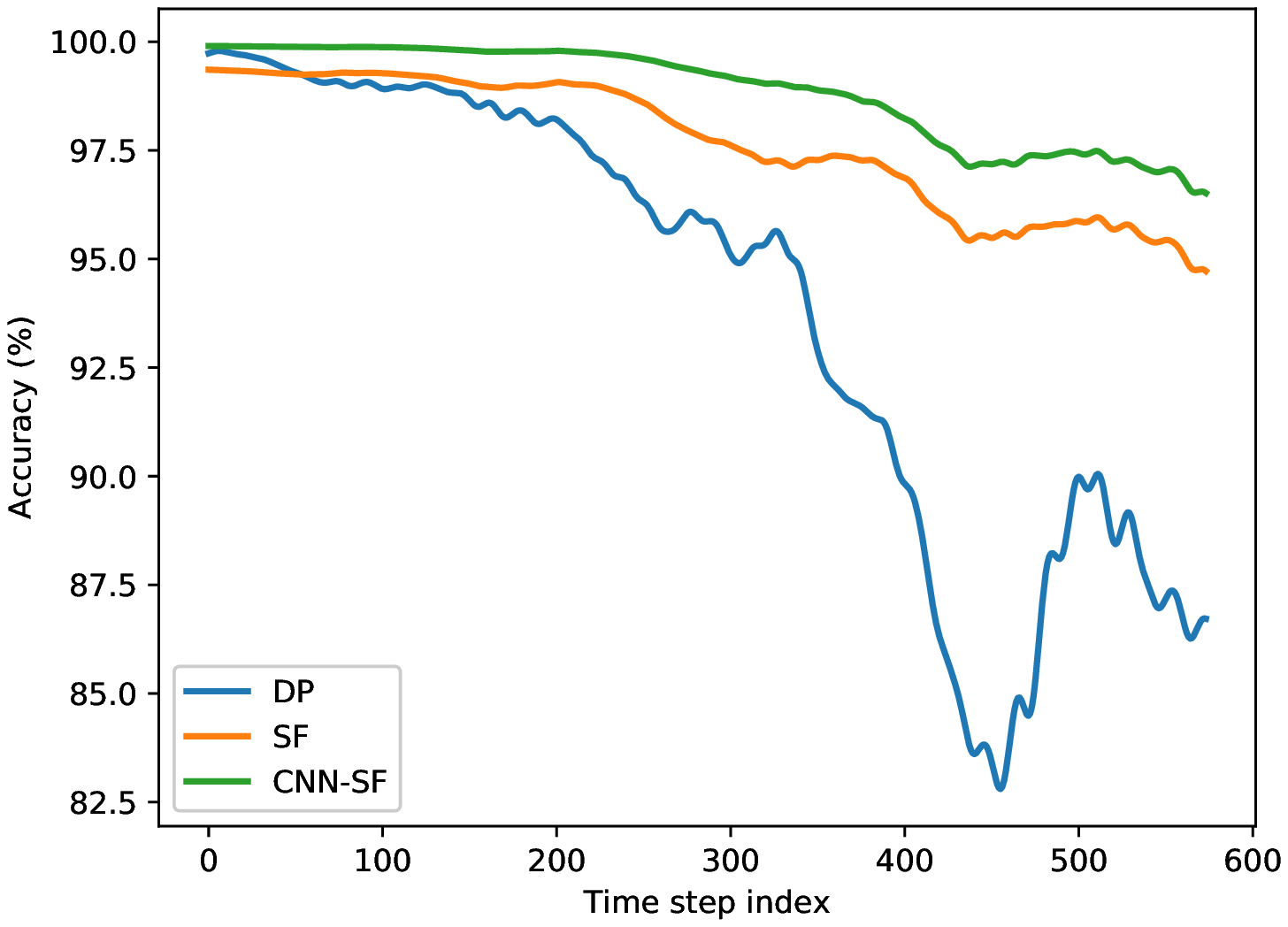}
  \caption{
Decomposition accuracy of DP, SF and CNN-SF algorithms of wavefield snapshots at different extrapolation time steps.
}
  \label{fig:accuracy.eps}
\end{figure}

To have a comprehensive analysis of the decomposition accuracy, 575 snapshots of coupled wavefields at different time steps are captured from the wavefield simulation in the Marmousi-2 model, and we use DP, SF and CNN-SF to generate the decomposed P-waves separately from all the snapshots. The accuracy of the three algorithms with increasing time steps are plotted in Figure~\ref{fig:accuracy.eps}. As the wave propagates through the model, more converted S-waves are generated, and the accuracies of all three algorithms decrease.
 Because the Marmousi-2 model has more velocity interfaces than the two-layer model, the accuracy of DP decreases significantly as the wave propagates deeper into the model. CNN-SF has the best accuracy among the three algorithms at all time steps.

\subsection{Application to elastic reverse-time migration}
In elastic reverse-time migration \cite{chang94}, the source wavefield and receiver wavefield are extrapolated forward and backward in time, respectively. Both the source and receiver wavefields are decomposed into P- and S-wave components, and various vector-based imaging conditions \cite{wenlong_vct15,zhao17,wenlong19} can be applied to generate images.
We apply the 2D dot-product imaging condition
\begin{equation}
I_{PP} = \sum_{t=0}^{T}\Big(U_{src\{x\}}^P U_{rec\{x\}}^P + U_{src\{z\}}^P U_{rec\{z\}}^P \Big)
\label{eqn:pp_image}
\end{equation}
for the PP image and
\begin{equation}
I_{PS} = \sum_{t=0}^{T}\Big( U_{src\{x\}}^P U_{rec\{x\}}^S + U_{src\{z\}}^P U_{rec\{z\}}^S \Big)
\label{eqn:ps_image}
\end{equation}
for the PS image, where $T$ is the number of discrete time steps during wavefield extrapolation; $U_{src}^P$, $U_{rec}^P$ and $U_{rec}^S$ are the decomposed P-waves from the source wavefield, and decomposed P- and S-waves from the receiver wavefield, respectively. All the wavefields in equations~\ref{eqn:pp_image} and \ref{eqn:ps_image} are understood to be functions of spatial position and time $t$; these variables are omitted for simplicity in all the wavefield expressions.
Elastic RTMs are performed with DP and CNN-SF for PS wavefield decomposition.

\begin{figure}[!t]
  \centering
  \includegraphics[width=3.2in]{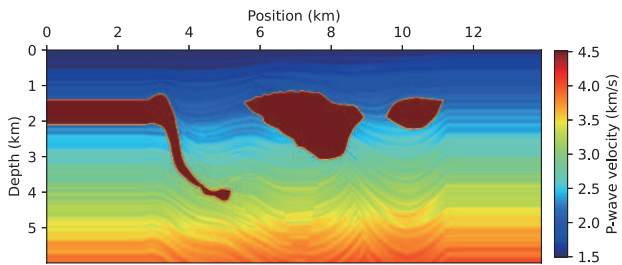}
  \caption{
The Pluto P-wave velocity model. The S-wave velocity model is approximated by multiplying the P-wave values by 0.6 at each grid point, and the density has a constant value of 2.1 $\mathrm{g/cm^3}$.
}
  \label{fig:model_pluto.eps}
\end{figure}

\begin{figure*}[!t]
  \centering
  \includegraphics[width=5.2in]{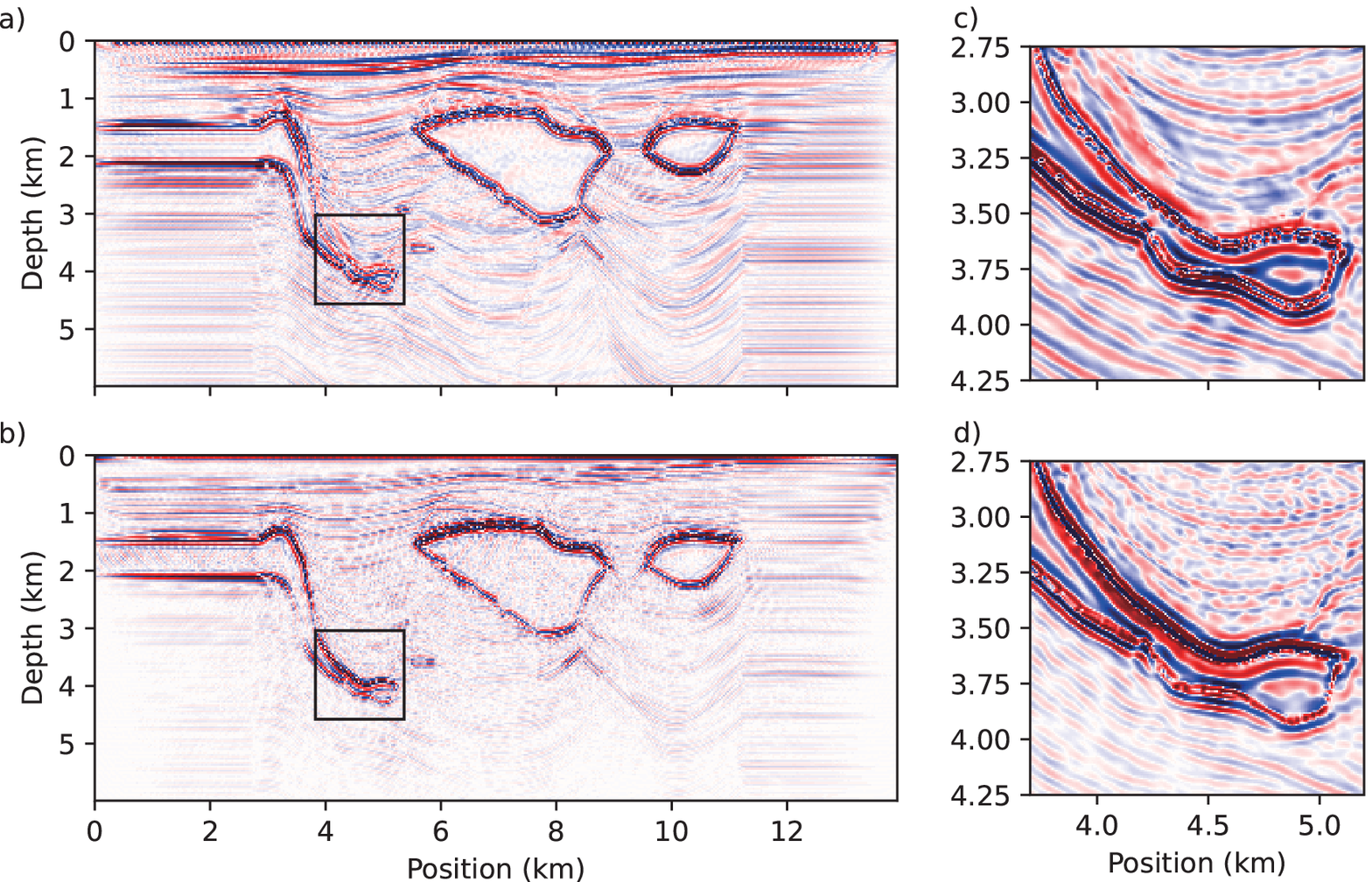}
  \caption{
The (a) PP and (b) PS images of elastic RTM using DP for PS decomposition; (c) and (d) are zoomed-in sections of boxed areas in (a) and (b), respectively. Artifacts that are generated by DP remain in the stacked images.
}
  \label{fig:image_sigsbee_dp.eps}
\end{figure*}

\begin{figure*}[!t]
  \centering
  \includegraphics[width=5.2in]{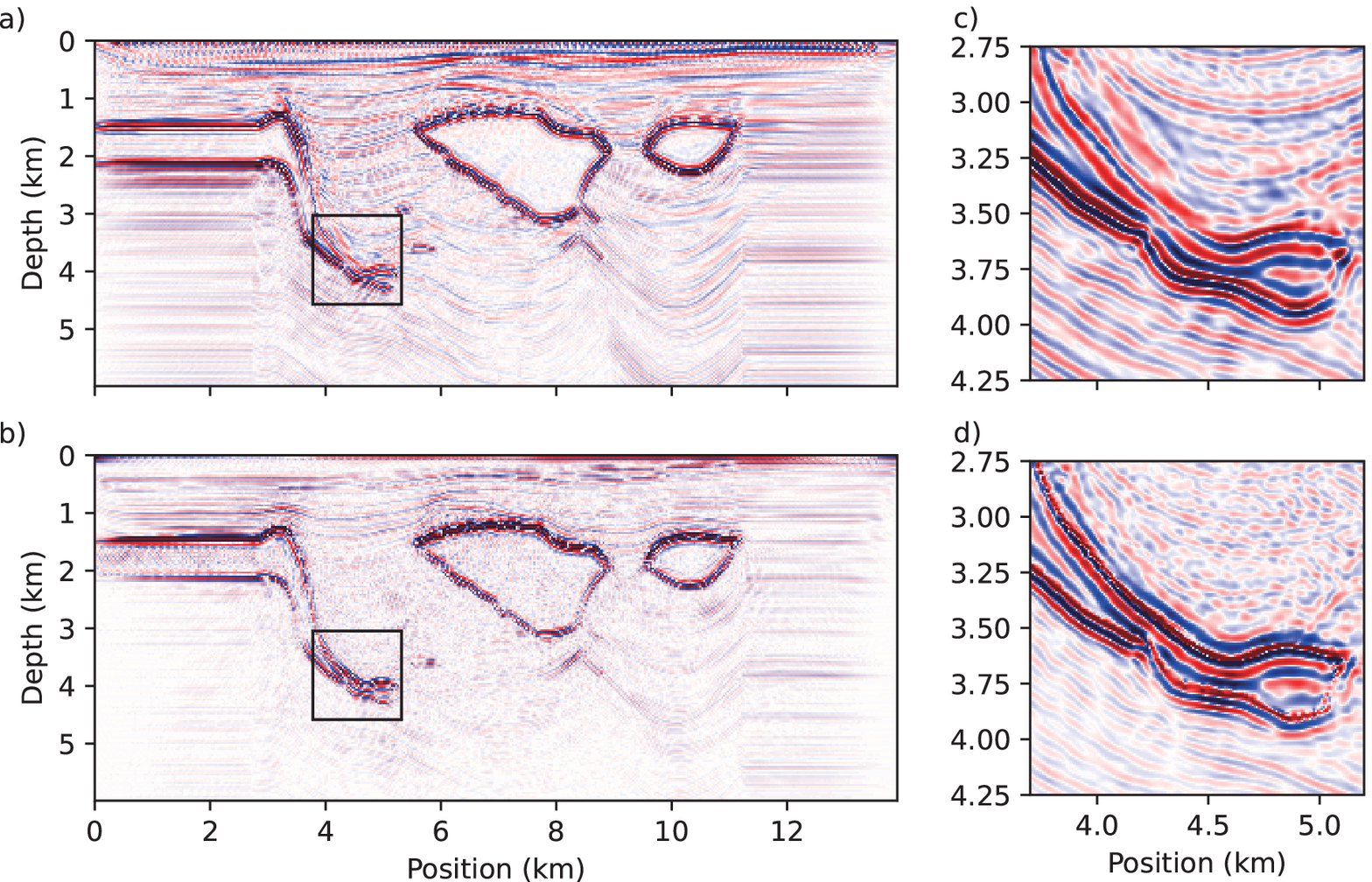}
  \caption{
The (a) PP and (b) PS images of elastic RTM using CNN-SF for PS decomposition; (c) and (d) are zoomed-in sections of boxed areas in (a) and (b), respectively. Compared with Figure~\ref{fig:image_sigsbee_dp.eps}, the artifacts are much reduced.
}
  \label{fig:image_sigsbee_cnn.eps}
\end{figure*}

The Pluto model \cite{stoughton01} is used in this test (Figure~\ref{fig:model_pluto.eps}).
The original Pluto model has only the P-wave velocity defined. We approximate the S-wave velocity model by multiplying the P-wave values by 0.6 at each grid point, and the density has a constant value of 2.1 $\mathrm{g/cm^3}$.
The grid increments are \textit{h} = 10 m in both spatial dimensions and time increment \textit{dt} = 1 ms. One hundred explosive sources with 15 Hz dominant frequency are initiated sequentially from (x, z) = (0.5, 0.0) km to  (13.5, 0.0) km with 0.13 km separation in the x direction. 1361 receivers are placed from (x, z) = (0.0, 0.0) to (13.6, 0.0) km with 0.01 km separation. The stacked PP and PS images using DP for PS decomposition are shown in Figure~\ref{fig:image_sigsbee_dp.eps}a and \ref{fig:image_sigsbee_dp.eps}b, respectively. The PS separation artifacts are visible along the high velocity contrasts in the zoomed-in sections of Figure~\ref{fig:image_sigsbee_dp.eps}c and d. Compare with Figures~\ref{fig:image_sigsbee_cnn.eps}a-\ref{fig:image_sigsbee_cnn.eps}d, which are the PP, PS and zoomed-in images using CNN-SF for PS decomposition; the image boundaries are more coherent than in Figure~\ref{fig:image_sigsbee_dp.eps} as the result of applying CNN-SF for PS decomposition.

\section{Discussion}
In the above tests, the dimensions of the CNN-tuned filters is fixed to be 15 $\times$ 15. The accuracy of decomposition can be further improved by using larger filters at the cost of computation efficiency.
The ground truth snapshots for training the CNN are the P-waves decomposed in the wavenumber domain, thus the accuracy of CNN-tuned filters can never surpass that of the wavenumber domain algorithm.

The computational complexity for PS decomposition using the 15 $\times$ 15 spatial filter set is $225N$, while the cost for the wavenumber domain algorithm is $4N$log$(N)$, thus the latter is also much more efficient than the former for small models. However, wavenumber domain algorithms require Fourier transforms which consume large RAM and have difficulty being applied locally in the model, while the spatial filters are HPC-friendly and can be easily applied in parallel using multiple cores. Spatial filters are suitable for processing target-oriented data sets, where PS decomposition can be performed locally on a portion of the wavefield and thus further reducing the cost.

The idea of using CNN-tuned spatial filters for PS decomposition can be extended to anisotropic media. In that case, the spatial filters are non-stationary across different positions within the model, because in anisotropic wavefields, the polarization directions of P- and S-waves depend on the model parameters \cite{yan09}, and several sets of filters can be tuned separately by training with parameters from representative points in the model. The proposed CNN-tuned spatial filters can also be extended to 3D, for which the number of filters increases to 6 ($L_x$, $L_y$, $L_z$, $L_{xy}$, $L_{xz}$ and $L_{yz}$) which need to be tuned as 3D convolutional operators.

\section{Conclusions}
We propose network-trained spatial filters for P- and S-wave decomposition in isotropic elastic wavefields.
The accuracy of decomposition using spatial filters is improved after CNN-tuning with no extra computation costs.
Migration tests with synthetic data show that the proposed spatial filters do not generate artifacts along high velocity contrasts as in the decoupled propagation method.
CNN-tuned spatial filters are suitable for parallel computation and target-oriented imaging.


\section*{Acknowledgment}
The research leading to this paper is supported by the National Natural Science Foundation of China (grant number: NSFC 41804108).
The participation of G.M. was supported by the UT-Dallas Geophysical Consortium.
\newpage
\bibliographystyle{IEEEtran}
\bibliography{ieee-twocolumn}

\clearpage

\clearpage

\end{spacing}
\end{document}